\documentclass{article}

\usepackage{arxivsimplified}

\usepackage[utf8]{inputenc} 
\usepackage[T1]{fontenc}    
\usepackage{hyperref}       
\usepackage{url}            
\usepackage{booktabs}       
\usepackage{amsfonts}       
\usepackage{nicefrac}       
\usepackage{microtype}      
\usepackage{lipsum}

\usepackage{graphicx}
\usepackage{subfigure}
\usepackage{cite}
\usepackage[dvipsnames]{xcolor}
\usepackage{amsmath}
\usepackage{blindtext}
\usepackage{amsfonts}
\usepackage{multirow}
\usepackage{array}
\usepackage{mathtools}
\usepackage{listings}
\usepackage{xcolor}
\definecolor{codegreen}{rgb}{0,0.6,0}
\definecolor{codegray}{rgb}{0.5,0.5,0.5}
\definecolor{codepurple}{rgb}{0.58,0,0.82}
\definecolor{backcolour}{rgb}{0.95,0.95,0.92}
\usepackage{esvect}
\usepackage{tabularx}
\newcolumntype{L}[1]{>{\raggedright\arraybackslash}p{#1}}
\newcolumntype{C}[1]{>{\centering\arraybackslash}p{#1}}
\newcolumntype{R}[1]{>{\raggedleft\arraybackslash}p{#1}}

\lstdefinestyle{mystyle}{
    backgroundcolor=\color{backcolour},   
    commentstyle=\color{codegreen},
    keywordstyle=\color{magenta},
    numberstyle=\tiny\color{codegray},
    stringstyle=\color{codepurple},
    basicstyle=\ttfamily\footnotesize,
    breakatwhitespace=false,         
    breaklines=true,                 
    captionpos=b,                    
    keepspaces=true,                 
    numbers=left,                    
    numbersep=5pt,                  
    showspaces=false,                
    showstringspaces=false,
    showtabs=false,                  
    tabsize=2
}
 
\usepackage{scalerel,stackengine}
\stackMath
\newcommand\reallywidehat[1]{%
\savestack{\tmpbox}{\stretchto{%
  \scaleto{%
    \scalerel*[\widthof{\ensuremath{#1}}]{\kern-.6pt\bigwedge\kern-.6pt}%
    {\rule[-\textheight/2]{1ex}{\textheight}}
  }{\textheight}%
}{0.5ex}}%
\stackon[1pt]{#1}{\tmpbox}%
}

\title{Data-driven recovery of hidden physics in reduced order modeling of fluid flows}

\author{
  Suraj Pawar  \\
  School of Mechanical \& Aerospace Engineering,\\
  Oklahoma State University, \\
  Stillwater, Oklahoma - 74078, USA.\\
  \texttt{supawar@okstate.edu} \\
   \And
   Shady E. Ahmed  \\
  School of Mechanical \& Aerospace Engineering,\\
  Oklahoma State University, \\
  Stillwater, Oklahoma - 74078, USA.\\
  \texttt{shady.ahmed@okstate.edu } \\
  \And
 Omer San \\
 School of Mechanical \& Aerospace Engineering,\\
  Oklahoma State University, \\
  Stillwater, Oklahoma - 74078, USA.\\
  \texttt{osan@okstate.edu} \\
  \And
  Adil Rasheed  \\
  Department of Engineering Cybernetics,\\
  Norwegian University of Science and Technology,\\
  N-7465, Trondheim, Norway.\\
  \texttt{adil.rasheed@ntnu.no}\\
}

\begin{document}
\maketitle

\begin{abstract}
In this article, we introduce a modular hybrid analysis and modeling (HAM) approach to account for hidden physics in reduced order modeling (ROM) of parameterized systems relevant to fluid dynamics. The hybrid ROM framework is based on using the first principles to model the known physics in conjunction with utilizing the data-driven machine learning tools to model remaining residual that is hidden in data. This framework employs proper orthogonal decomposition as a compression tool to construct orthonormal bases and Galerkin projection (GP) as a model to built the dynamical core of the system. Our proposed methodology hence compensates structural or epistemic uncertainties in models and utilizes the observed data snapshots to compute true modal coefficients spanned by these bases. The GP model is then corrected at every time step with a data-driven rectification using a long short-term memory (LSTM) neural network architecture to incorporate hidden physics. A Grassmannian manifold approach is also adapted for interpolating basis functions to unseen parametric conditions. The control parameter governing the system's behavior is thus implicitly considered through true modal coefficients as input features to the LSTM network. The effectiveness of the HAM approach is discussed through illustrative examples that are generated synthetically to take hidden physics into account. Our approach thus provides insights addressing a fundamental limitation of the physics-based models when the governing equations are incomplete to represent underlying physical processes. 
\end{abstract}

\keywords{Reduced order modeling \and Deep learning \and Neural networks \and Parameterized systems \and Proper orthogonal decomposition \and Grassmannian manifold \and Long short-term memory embedding \and Hidden physics \and Structural uncertainty}

\section{Introduction}
Advances in machine learning algorithms along with the huge amount of data generated from high-fidelity numerical simulations, lab experiments, sensors data can be integrated with physical modeling to improve the prediction of complex dynamical systems. The physics-based approaches are interpretable and can be extrapolated beyond the observed data, while machine learning algorithms can discover unknown patterns from the data. Both of these approaches are complementary and can benefit from one another. Readers are directed to \cite{reichstein2019deep} and references therein for an excellent perspective on the hybridization of physical modeling and machine learning  algorithms in the context of geoscientific research. 

The problems in fluid mechanics are very high-dimensional owing to a wide range of spatial and temporal scales that have to be resolved which places an enormous computational burden on numerical simulations. There is a range of techniques that aim at constructing a reduced-order model (ROM) which captures the essential features of these flows and is also computationally orders of magnitudes faster than actual numerical simulations \cite{rowley2017model,taira2019modal}. Furthermore, many physical processes are governed by parameterized partial differential equations and there is a growing interest in building parametric ROMs with high-fidelity over a range of parameters \cite{benner2015survey}. These ROMs are particularly essential for system identification, design optimization, flow control, and uncertainty quantification that require multiple forward simulations over a large range of parameters. Proper orthogonal decomposition (POD) is one of the most popular model reduction methods which decomposes the flow field into a set of basis functions that optimally describes the system and selects only the most energy-conserving bases to represent the system \cite{sirovich1987turbulence}. The POD is complemented by a Galerkin projection (GP) in which the dynamics of the system is modeled. 
Recently, there is a significant effort to use machine learning algorithms in model order reduction of nonlinear fluid flows \cite{lee2018model, brunton2019machine, san2018neural, xiao2019domain, swischuk2019projection}.

Often there is a discrepancy between the physical model and the observed data. This discrepancy arises due approximations to complex physical processes, the requirement of parameterizations in physical models, imperfect knowledge about source terms, observational errors, etc. Machine learning algorithms have been utilized in identifying and extracting patterns from the observational data that can help us in the understanding of the physical phenomena and discovering the equations governing these phenomena \cite{schmidt2009distilling, rudy2017data, raissi2018hidden}. Similarly, machine learning can be used to extract patterns from the observed data that are not included in a physical model and will allow us to improve the physical model. The goal of the present study is to build ROM for any parametrized process that can be defined by
\begin{equation}
    \frac{\partial {\mathbf{u}}}{\partial t}({\mathbf{x}},t;\mu. \kappa)={\mathbf{F}_m}({\mathbf{x}},t;{\mathbf{u}};\mu) + {\mathbf{\Pi}}({\mathbf{x}},t;{\mathbf{u}};\mu,\kappa),
    \label{eq:physical_system}
\end{equation} 
where ${\mathbf{u}}$ is the prognostic variable, ${\mathbf{F}_m}$ is the physics-based model (i.e., dynamical core) governing the known processes, and ${\mathbf{\Pi}}$ includes the external empirical parameterizations and unknown physics. Here, $\mu$ and $\kappa$ refer to the control parameters that are relevant to the dynamical core and the hidden physics, respectively. The source term ${\mathbf{\Pi}}$ is usually unknown and can be learned using machine learning methods from the observed data. However, the machine learning methods and in particular deep learning models lack interpretability and are prone to produce physically inconsistent results. Hence, there is an active research going on to incorporate the physical models in machine learning methods to make them physically consistent, such as loss regularization based on physical laws \cite{raissi2019physics}, designing novel neural network to embed certain physical properties \cite{ling2016reynolds}, and building hybrid models to correct the imperfect knowledge in physical models \cite{pan2019physics,mou2019data}. How should we inject physics and domain knowledge into machine learning models? When and how machine learning might be a complementary approach for generating more robust surrogate models? These are open research questions that we tackle, and the present study offers a glimpse to addressing these issues and pertains to introducing an effective hybrid modeling approach toward more accurate real-time predictions of fluid flows under epistemic or structural uncertainties.   

The motivation behind this work is to synthesize the physics-based approach  along with machine learning to model the hidden physics in parameterized ROMs. The paper is structured as follow. We discuss the formulation of hybrid analysis and modeling (HAM) framework in Section~\ref{sec:ham}. The numerical results for the one-dimensional Burgers equation and two-dimensional Navier-Stokes equations with HAM framework are detailed in Section~\ref{sec:results}. Finally Section~\ref{sec:conclusion} will present conclusions and ideas for the future work.

\section{Hybrid analysis and modeling (HAM) framework} \label{sec:ham}
To construct a set of orthonormal POD basis functions, we collect the data snapshots, $\mathbf{u}_1, \mathbf{u}_2, \dots, \mathbf{u}_N$ $\in$ $\mathbb{R}^m$, at different time instants. We form the matrix $\mathbf{A}$ $\in$ $\mathbb{R}^{m \times N}$ whose columns are the snapshots $\mathbf{u}_n$, and then perform the reduced singular value decomposition (SVD) of the matrix
\begin{equation}
    \mathbf{A} = \mathbf{W}\mathbf{\Sigma}\mathbf{V}^T = \sum_{k=1}^r\sigma_k \mathbf{w}_k \mathbf{v}_k^T,
\end{equation}
where $r$ is the rank of $\mathbf{A}$, $\mathbf{W}$ is an $m \times r$ matrix with orthonormal columns $\mathbf{w}_k$, $\mathbf{V}$ is an $N \times r$ matrix with orthonormal columns $\mathbf{v}_k$, and $\mathbf{\Sigma}$ is an $r \times r$ matrix with positive diagonal entries, called singular values, arranged such that $\sigma_1 \ge \sigma_2 \ge \dots \ge \sigma_r > 0$. The vectors $\mathbf{w}_k$ are the POD modes which we denote as $\phi_k$ in this text, and $\mathbf{\Phi}=\{ \phi_k\}_{k=1}^{R}$ is the set of POD basis functions for any values of $R \leq r$ \cite{rowley2017model}. The representation of the approximated field using the POD modes is as follow,
\begin{equation} \label{eq:upod}
    \mathbf{u}(\mathbf{x}, t_n) = \sum_{k=1}^{R} a_k^{(n)}(t_n) \phi_k(\mathbf{x}),
\end{equation}
where $a_k^{(n)}$ are the time dependent modal coefficients. The GP equations are obtained by applying projection to our physical system (i.e., using the linear superposition given by Equation~\ref{eq:upod} in Equation~\ref{eq:physical_system} and applying inner product of the resulting equation with the basis functions $\phi_k$ that are orthonormal to each other). The resulting system of equations is given below 
\begin{equation}
    \dfrac{\text{d}a_k}{\text{d}t} =  \underbrace{\sum_{i=1}^{R} \mathfrak{L}_{ik} a_i + \sum_{i=1}^{R} \sum_{j=1}^{R} \mathfrak{N}_{ijk} a_i a_j}_{\text{Physics-based model}~{G(a_k)}} + \underbrace{{\tilde{C}}_{k}}_{\text{Hidden physics}},
\end{equation}
where $\mathfrak{L}$ and $\mathfrak{N}$ are the linear and nonlinear operator of the physical system, and $\tilde{C}$ is the part of the system for which the physical model is not available. The linear and nonlinear operator for two prototypical examples investigated in this study are presented in Table~\ref{tab:galerkin_operators}. The angle-parentheses in Table~\ref{tab:galerkin_operators} refers to the Euclidean inner product defined as $\langle \mathbf{x} , \mathbf{y} \rangle = \mathbf{x}^T \mathbf{y} =  \sum_{i=1}^{m} x_iy_i$. In a discrete sense, the update formula can be written
\begin{equation} \label{eq:ak}
    {a}_k^{(n+1)} = {a}_k^{(n)} + {\Delta t} \sum_{q=0}^{s} \beta_q [G({a}_k^{(n-q)}) + {\tilde{C}}_{k}^{(n-q)}]  ,
\end{equation}
where $s$ and $\beta_q$ depends upon the numerical scheme used for the time integration. We use the third-order Adams-Bashforth (AB3) method in this study for which $s=2,~ \beta_0=23/12,~ \beta_1=-16/12,$ and $\beta_2=5/12$. The physics-based GP model at any time $t_n$ is
\begin{equation}
   G({a}_k^{(n)}) = \sum_{i=1}^{R} \mathfrak{L}_{ik} a_{i}^{(n)} + \sum_{i=1}^{R} \sum_{j=1}^{R} \mathfrak{N}_{ijk} a_{i}^{(n)} a_{j}^{(n)}. 
\end{equation}
In our HAM approach, since we attempt learning the hidden physics part $\tilde{C}$ by using a supervised data-driven approach, Equation~\ref{eq:ak} can be rewritten as 
\begin{equation} \label{eq:depl}
    {a}_k^{(n+1)} = {a}_k^{(n)} + {\Delta t} \sum_{q=0}^{s} \beta_q G({a}_k^{(n-q)}) + {C}_{k}^{(n+1)}.
\end{equation}
We can generate training set from the true projection of observed data into POD basis. These true projection modal coefficients encompasses both the dynamical core of the system and hidden physics. The true projection modal coefficients are computed as below
\begin{equation} \label{eq:true}
    \alpha_k^{(n)} = \langle \mathbf{u}(\mathbf{x}, t_n) , \phi_k \rangle.
\end{equation}
Using Equation~\ref{eq:depl} and~\ref{eq:true} we can recover the hidden physics as below
\begin{equation}
    {C}_{k}^{(n+1)} = \alpha_k^{(n+1)} - \bigg[ {\alpha}_k^{(n)} + {\Delta t} \sum_{q=0}^{s} \beta_q G({\alpha}_k^{(n-q)}) \bigg].
\end{equation}
We can use any of the suitable machine learning algorithm to learn this correction term $C_k$. In this study we employ LSTM neural network algorithm to learn the mapping from true modal coefficients to the correction term (i.e., $\{\alpha_1,\dots,\alpha_R\}\in \mathbb{R}^R \rightarrow \{C_1,\dots,C_R\} \in \mathbb{R}^R$). To be consistent with the underlying AB3 scheme, in this study we use lookback of 3 steps in our LSTM architecture (e.g., see \cite{rahman2019non} for further details). During the deployment, we start with an initial condition and then at every time step the correction is added to the GP model $G(a_k)$, and we proceed this process in a recursive fashion. The parameter governing the behavior of the physical system is thus taken implicitly into account through physics-based GP model.

\begin{table}
\begin{center}
\begin{tabular}{C{0.27\textwidth}C{0.3\textwidth}C{0.35\textwidth}} \hline \\ [-1.5ex]
Example    & $ \mathfrak{L}_{ik}$ & $\mathfrak{N}_{ijk} $ \vspace{2mm}  \\ \hline \\ [-1.5ex]  
Burgers equation & $\bigg \langle \dfrac{1}{\text{Re}} \dfrac{\partial^2 \phi_i^u}{\partial x^2} , \phi_k^u \bigg \rangle$ & $ \bigg \langle -\phi_i^u \dfrac{\partial \phi_j^u}{\partial x},\phi_k^u \bigg \rangle$ \vspace{1mm}\\
Navier-Stokes equations & $ \bigg \langle \dfrac{1}{\text{Re}} \bigg(\dfrac{\partial^2 \phi_i^{\omega}}{\partial x^2} + \dfrac{\partial^2 \phi_i^{\omega}}{\partial y^2}\bigg),\phi_k^{\omega} \bigg \rangle $ & $\bigg \langle -\bigg(\dfrac{\partial \phi_i^{\omega}}{\partial x} \dfrac{\partial \phi_j^{\psi}}{\partial y} -  \dfrac{\partial \phi_i^{\omega}}{\partial y} \dfrac{\partial \phi_j^{\psi}}{\partial x}\bigg),\phi_k^{\omega} \bigg \rangle$ \vspace{2mm}  \\ \hline  \vspace{2mm} 
\end{tabular}
\caption{Galerkin projection operators for two prototypical examples where $\phi_k^{u}$ refers to basis functions of $u$ field in the Burgers equation. Similarly, in the Navier-Stokes equations, $\phi_k^{\omega}$ and $\phi_k^{\psi}$ refer to basis functions of the vorticity and streamfunction, respectively.}
\label{tab:galerkin_operators}
\end{center}
\end{table}

\begin{figure}
\centering
\mbox{
	\subfigure{\includegraphics[width=0.95\textwidth]{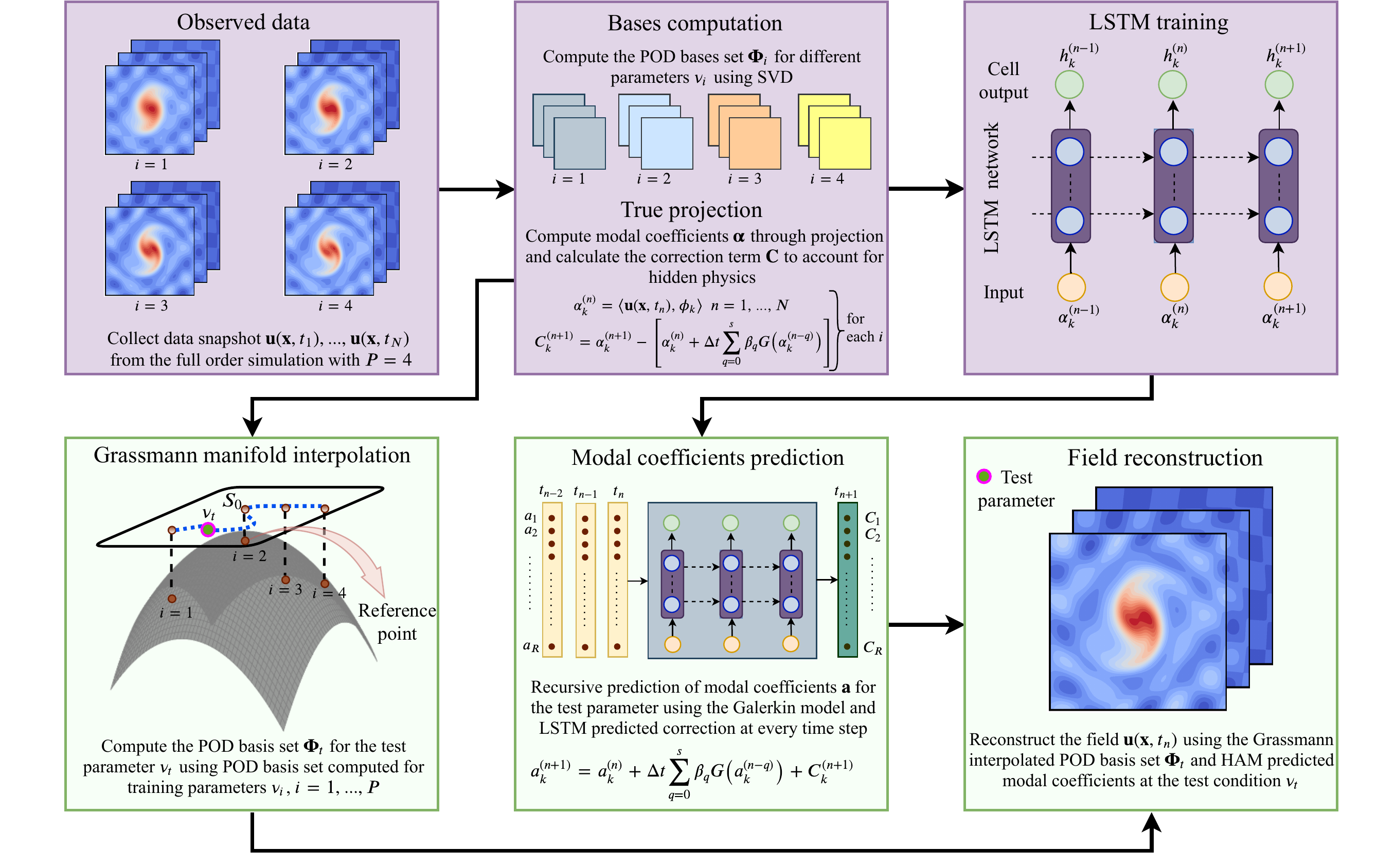}}}
\caption{Hybrid analysis and modeling (HAM) framework for model order reduction.} 
\label{fig:ham_rom}
\end{figure}

During training, we compute the POD basis sets $\mathbf{\Phi} = [\phi_1, \phi_2, ..., \phi_R]$ for different parameters governing the physical system. We utilize a Grassmann manifold interpolation approach \cite{amsallem2008interpolation, zimmermann2018geometric} to compute the POD basis set for the test parameter from these POD basis sets during deployment. The Grassman manifold interpolation consists of choosing a reference point $S_{0}$ corresponding to the basis $\mathbf{\Phi}_0$, and then mapping each point $S_i$ to a matrix $\mathbf{\Gamma}_i$ which represents the tangent space $S_{0}$ using logarithm map $\text{Log}_{S_{0}}$
\begin{equation}
    (\mathbf{\Phi}_i-\mathbf{\Phi}_0 \mathbf{\Phi}_0^T \mathbf{\Phi}_i)(\mathbf{\Phi}_0^T\mathbf{\Phi}_i)^{-1} = \mathbf{U}_i\mathbf{\Sigma}_i \mathbf{V}_i^T,
\end{equation}
\begin{equation}
    \mathbf{\Gamma}_i =  \mathbf{U}_i \text{tan}^{-1}(\mathbf{\Sigma}_i) \mathbf{V}_i^T.
\end{equation}
The matrix $\mathbf{\Gamma}_t$ corresponding to the test parameter $\nu_t$ is obtained by the Lagrange interpolation of the matrices $\mathbf{\Gamma}_i$ corresponding to $\nu_i$, $i=1,2, \dots, P$,  
\begin{equation}
    \mathbf{\Gamma}_t = \sum_{i=1}^{P}\bigg( \prod_{\substack{j=1 \\ j\neq i}}^{P}\frac{\nu_t - \nu_j}{\nu_i - \nu_j}\bigg)\mathbf{\Gamma}_i,
\end{equation}
where $P$ refers to the number of the control parameters for offline simulations (e.g., $P=4$ in our numerical examples).
The POD basis $\mathbf{\Phi}_t$ corresponding to the test parameter $\nu_t$ is computed using the exponential mapping as follow 
\begin{equation}
    \mathbf{\Gamma}_t = \mathbf{U}_t \mathbf{\Sigma}_t \mathbf{V}_{t}^{T},
\end{equation}
\begin{equation}
    \mathbf{\Phi}_t = [\mathbf{\Phi}_0 \mathbf{V}_t \text{cos}(\mathbf{\Sigma}_t) + \mathbf{U}_t \text{sin}(\mathbf{\Sigma}_t)]\mathbf{V}_{t}^{T}.
\end{equation}
We note that trigonometric functions apply only to diagonal elements. The main blocks of the hybrid analysis and modeling (HAM) framework are shown in Figure~\ref{fig:ham_rom}. We use two hidden layers with 80 cells for both Burgers equation and Navier-Stokes equations to train the LSTM network. Our experiments with different set of hyperparameters show that the LSTM network is not highly sensitive to hyperparameters. 

\section{Numerical experiments} \label{sec:results}
\subsection{One-dimensional Burgers equation}
We test the performance of HAM framework for one-dimensional Burgers equation which is a prototypical example for nonlinear advection-diffusion problems. The Burgers equation can be written as 
\begin{equation}
    \dfrac{\partial u}{\partial t} + u \dfrac{\partial u}{\partial x} = \dfrac{1}{\text{Re}} \dfrac{\partial ^2 u}{\partial x^2} + \Pi, \quad x \in [0,1], \quad t \in [0,1.5]
    \label{eq:burgers}
\end{equation}
and accepts an analytical solution in the form of 
\begin{equation}
    \bar{u}(x,t) = \frac{\frac{x}{t+1}}{1 + \sqrt{\frac{t+1}{t_0}} \text{exp} \big(\text{Re} \frac{x^2}{4t + 4} \big)},
\end{equation}
when the hidden physics part is set to $\Pi=0$.
Here, $t_0=\text{exp}(\text{Re}/8)$ and Re is the Reynolds number that parameterizes Burgers equation. We add 30\% perturbation to this canonical solution to represent the hidden physics with unknown source term. Therefore, our snapshots read from 
\begin{equation}  \label{eq:bur}
    u(x,t) = 1.3\frac{\frac{x}{t+1}}{1 + \sqrt{\frac{t+1}{t_0}} \text{exp} \big(\text{Re} \frac{x^2}{4t + 4} \big)},
\end{equation}
and we note that $\Pi \neq0$. In our HAM approach, we model this part $\Pi$ with a data-driven machine learning model. We generate the data snapshots using the analytical solutions given by Equation~\ref{eq:bur} for $\text{Re} = [200,400,600,800]$ and evaluate the performance of HAM framework for $\text{Re}=500$, and $1000$. The reference point in the Grassmannian manifold interpolation is set to $\text{Re}=400$ and $\text{Re}=800$ for test $\text{Re}=500$, and $1000$, respectively.
The performance of HAM framework for these test Reynolds numbers can inform us about its interpolatory and extrapolatory prediction capability.

\begin{figure}
\centering
\mbox{
	\subfigure{\includegraphics[width=0.95\textwidth]{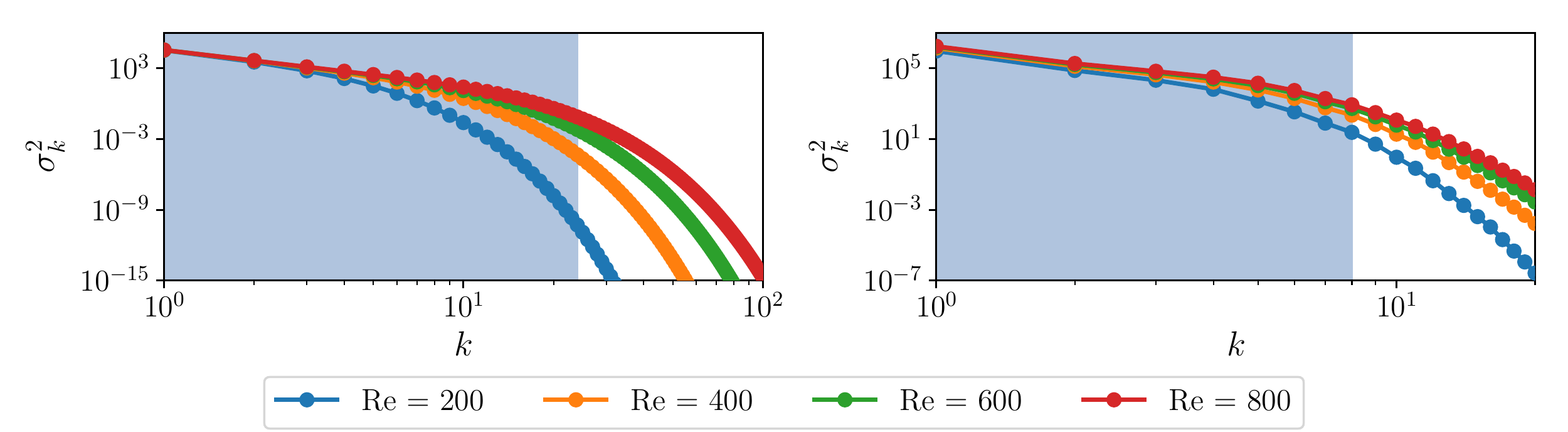}}}
\caption{Square of singular values of the snapshot data matrix $\mathbf{A}$ (equivalent to eigenvalues of $\mathbf{A}\mathbf{A}^T$ or $\mathbf{A}^T\mathbf{A}$) for different Re training data sets obtained by the 1D Burgers equation (left) and the 2D Navier-Stokes equations with $\gamma=0.1$ (right). } 
\label{fig:eigen_history}
\end{figure}

\begin{figure}
\centering
\mbox{
	\subfigure{\includegraphics[width=0.95\textwidth]{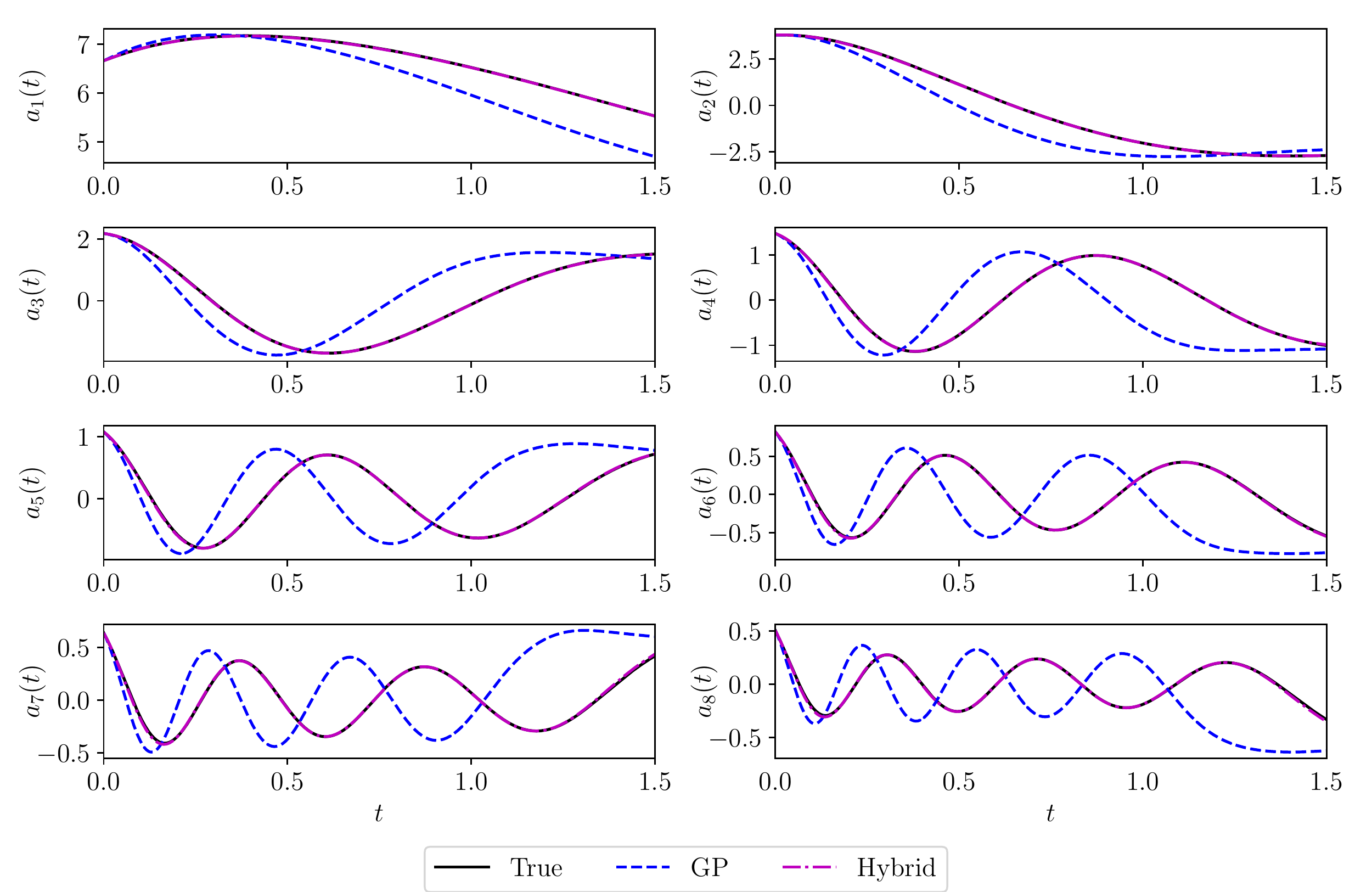}}
}
\caption{Temporal evolution of modal coefficients for the Burgers equation at $\text{Re}=500$.} 
\label{fig:burgers_modes_500}
\end{figure}

\begin{figure}
\centering
\mbox{
	\subfigure{\includegraphics[width=0.95\textwidth]{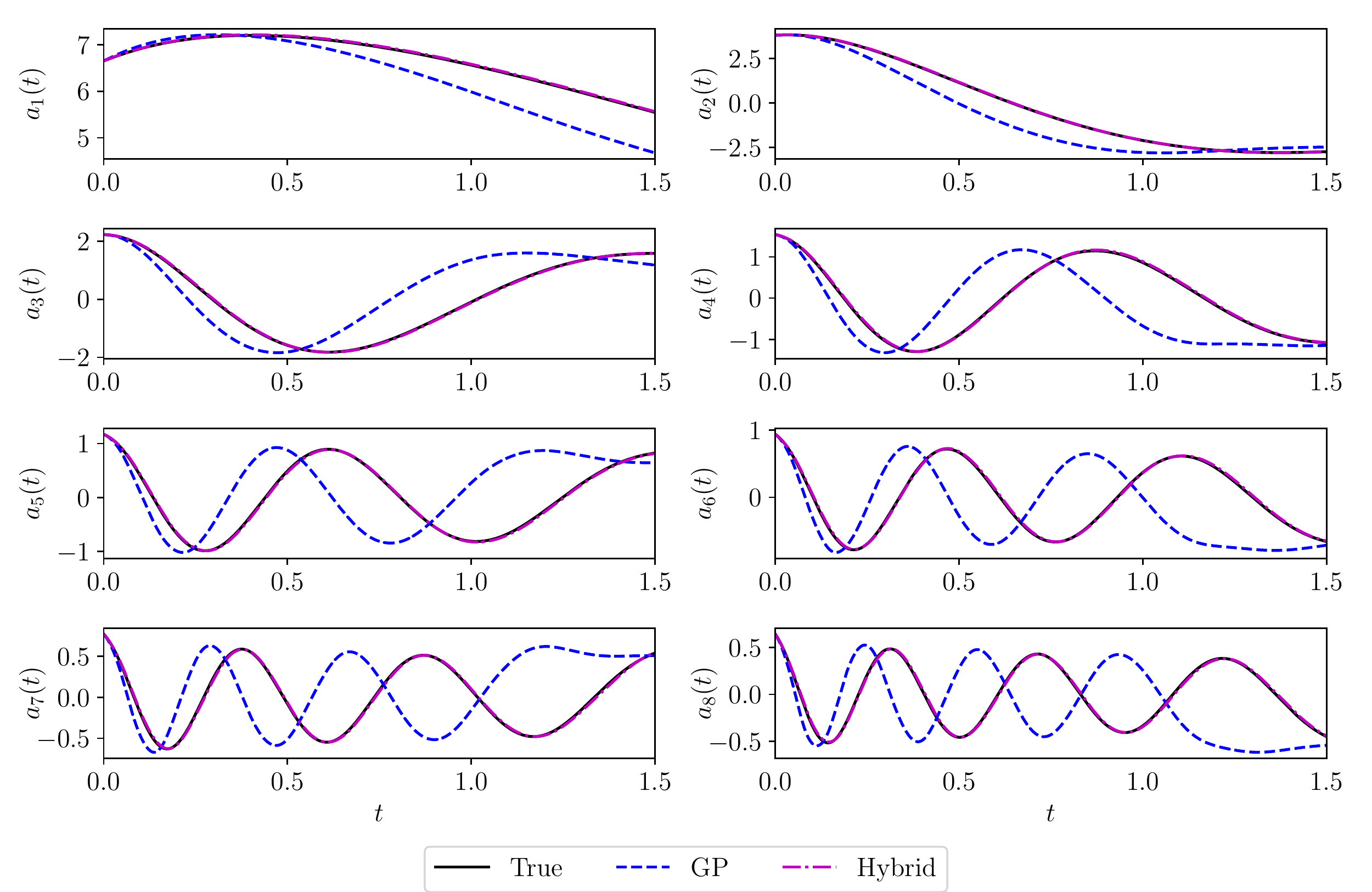}}
}
\caption{Temporal evolution of modal coefficients for the Burgers equation at $\text{Re}=1000$.}
\label{fig:burgers_modes_1000}
\end{figure}

\begin{figure}
\centering
\mbox{
	\subfigure{\includegraphics[width=0.95\textwidth]{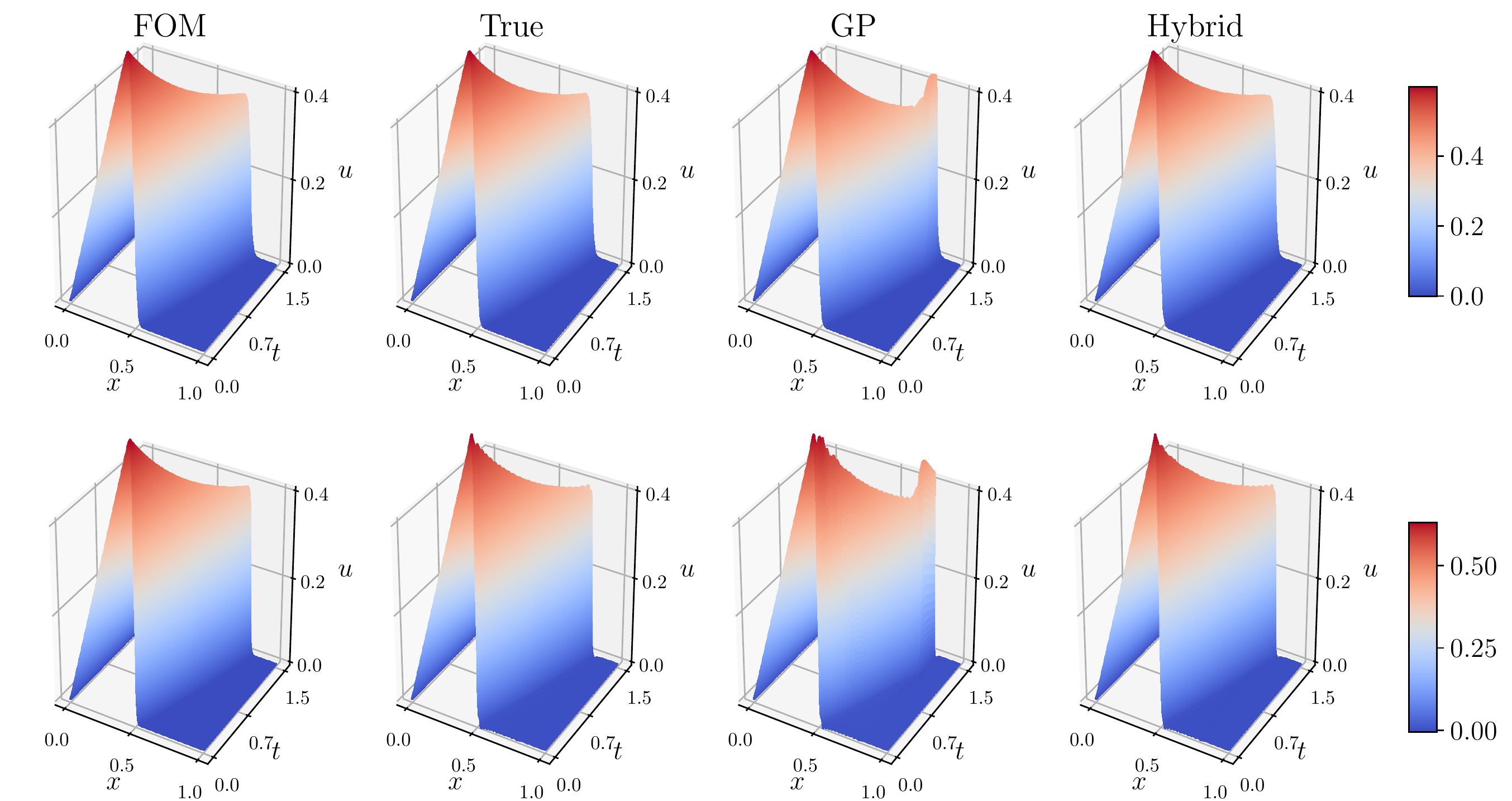}}
}
\caption{Space-time solution field for the Burgers equation for an interpolatory test condition at $\text{Re}=500$ (top), and an extrapolatory test condition at $\text{Re}=1000$ (bottom).}
\label{fig:burgers_field}
\end{figure}

We use $R=24$ modes in our ROM to represent the Burgers equation which captures more than 99.99\% energy at all Reynolds numbers. The eigenvalues of Burgers equation for different Reynolds number included in training is shown in Figure~\ref{fig:eigen_history}. 
From Figure~\ref{fig:burgers_modes_500} and ~\ref{fig:burgers_modes_1000}, we observe that in the presence of source term, the GP is not able to predict the same trajectory as true projection even with 24 modes (please note that we show only first 8 modes in the plots). The HAM framework adds the correction due to source term at every time step and follows the same trajectory similar to true projection modal coefficients as shown in Figure~\ref{fig:burgers_modes_500} for interpolatory Reynolds number $\text{Re}=500$. We see some discrepancy in true and hybrid modal coefficients for $\text{Re}=1000$ which can be attributed to the limited extrapolation capability of data-driven methods as demonstrated for turbulent isotropic flows \cite{erichson2019shallow}. Figure~\ref{fig:burgers_field} shows the time evolution of the solution field from $t=[0,1.5]$ for FOM and reconstructed solution field with true, GP, and hybrid modal coefficients. The HAM framework can produce the solution field similar to the true projection for both $\text{Re}=500$ and $1000$. We also notice that for $\text{Re}=1000$ more modes are needed in ROM to get the same level of accuracy as FOM. 

\subsection{Two-dimensional Navier-Stokes equations}
After a successful demonstration of the HAM framework for the Burgers equation, we test the HAM framework for two-dimensional Navier-Stokes equations applied to the vortex-merger problem. This problem is a prototypical example used to study the merging of two co-rotating vortices. Initially, the two vortices of the same sign are separated by some distance which end up as a single nearly axisymmetric vortex after some time. The two-dimensional Navier-Stokes equations in vorticity-streamfunction formulation can be written as
\begin{align} \label{eq:ns2d}
\dfrac{\partial \omega}{\partial t} + \dfrac{\partial \psi}{\partial y}\dfrac{\partial \omega}{\partial x} - \dfrac{\partial \psi}{\partial x} \dfrac{\partial \omega}{\partial y} &= \dfrac{1}{\text{Re}}\left( \dfrac{\partial^2 \omega}{\partial x^2} + \dfrac{\partial^2 \omega}{\partial y^2} \right) + \Pi,
\end{align}
\begin{align} \label{eq:ns2d-mass}
\dfrac{\partial^2 \psi}{\partial x^2} + \dfrac{\partial^2 \psi}{\partial y^2}=-\omega 
\end{align}
where $\omega$ is the vorticity defined as $\omega = \nabla \times \mathbf{u}$, $\mathbf{u} = [u,v]^T$ is the velocity vector, and $\psi$ is the streamfunction. We start from an initial vorticity field of two Gaussian-distributed vortices,
\begin{align}
    \omega(x, y, 0) = \exp\left( -\pi \left[ (x-x_1)^2  + (y-y_1)^2 \right] \right) + \exp{\left( -\pi \left[ (x-x_2)^2 + (y-y_2)^2 \right] \right)},
\end{align}
where their centers are initially located at $(x_1,y_1) = (3\pi/4,\pi)$ and $(x_2,y_2) = (5\pi/4,\pi)$. We add a perturbation field (i.e., referring to hidden physics) by utilizing an arbitrary array of decaying Taylor-Green vortices as the source term, which is given below
\begin{equation}
    \Pi = -F(t)\text{cos}(3 x) \text{cos}(3 y),
\end{equation}
where $F(t)= \gamma~ e^{-t/\text{Re}}$. We use computational domain $(x,y) \in [0,2\pi]$ with periodic boundary conditions. We generate data snapshots for $\text{Re}=[200,400,600,800]$ with $256^2$ spatial grid and a time-step of 0.01 from time $t=0$ to $t=20$. We test the HAM framework for out-of-sample conditions at $\text{Re}=500$ (interpolatory) and $\text{Re}=1000$ (extrapolatory). The Grassmannian manifold interpolation utilizes the data set from $\text{Re}=400$ and $\text{Re}=800$ as the reference set for test $\text{Re}=500$ and $\text{Re}=1000$, respectively.  

\begin{figure}
\centering
\mbox{
	\subfigure{\includegraphics[width=0.95\textwidth]{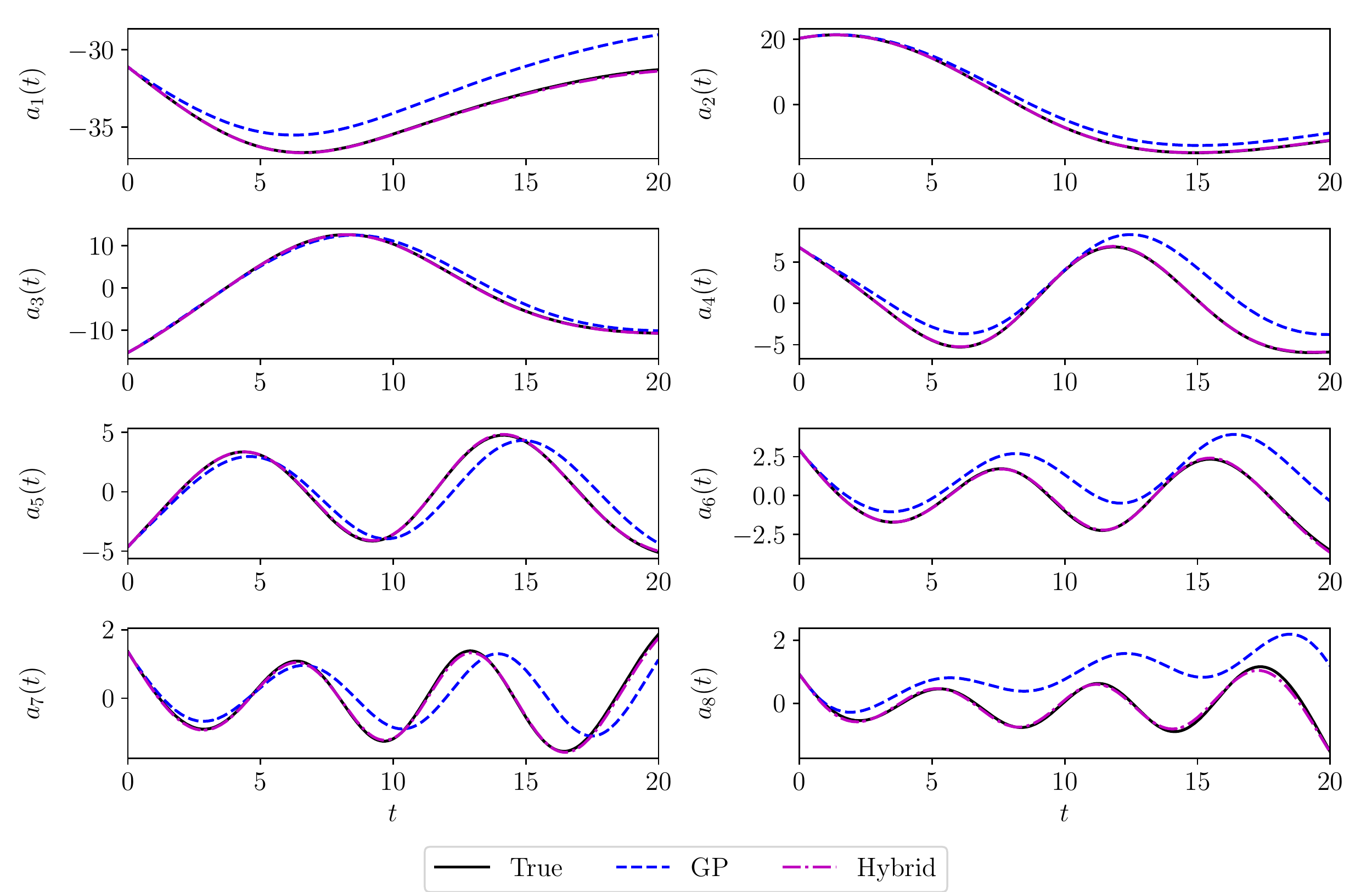}}
}
\caption{Temporal evolution of vorticity modal coefficients for the vortex-merger problem at $\text{Re}=500$ and $\gamma=0.01$.} 
\label{fig:ns_modes_500}
\end{figure}

\begin{figure}
\centering
\mbox{
	\subfigure{\includegraphics[width=0.95\textwidth]{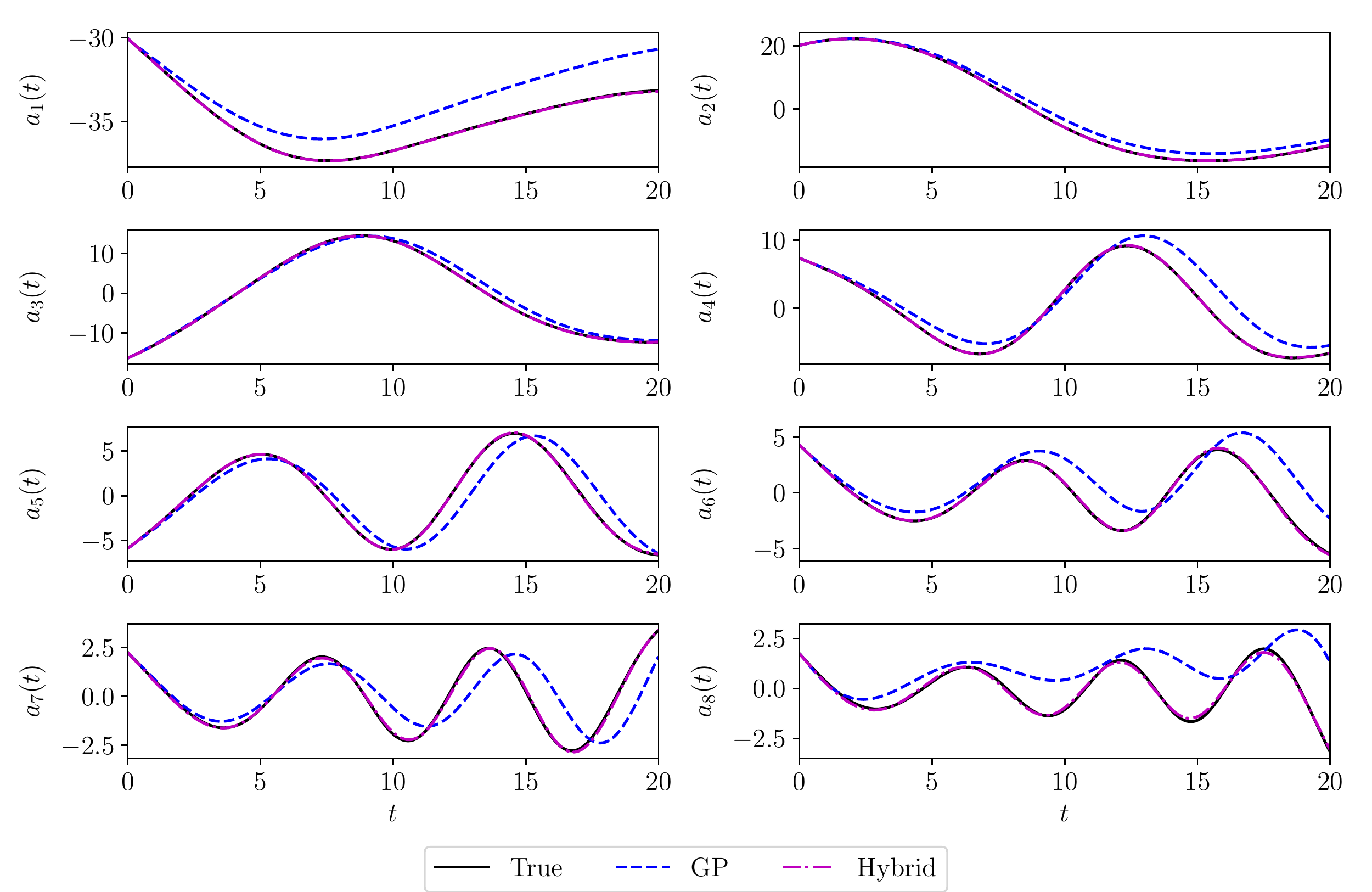}}
}
\caption{Temporal evolution of vorticity modal coefficients for the vortex-merger problem at $\text{Re}=1000$ and $\gamma=0.01$.} 
\label{fig:ns_modes_1000}
\end{figure}

\begin{figure}
\centering
\mbox{
	\subfigure{\includegraphics[width=1.0\textwidth]{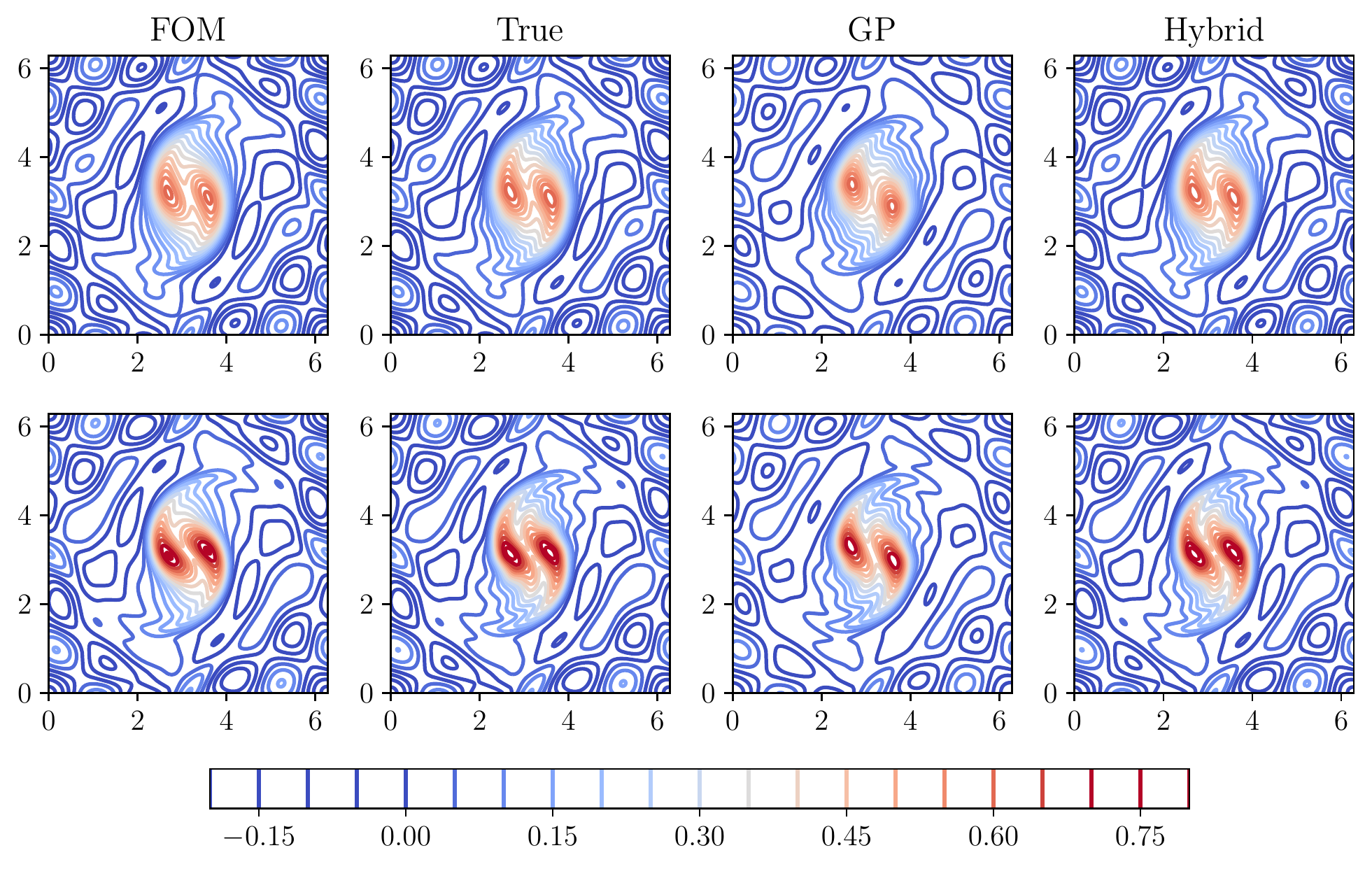}}
}
\caption{Vorticity field at time $t=20$ for the vortex-merger problem testing an interpolatory condition at $\text{Re}=500$ (top), and an extrapolatory condition at $\text{Re}=1000$ (bottom) with $\gamma=0.01$.} 
\label{fig:ns_field}
\end{figure}

We use 8 modes in our ROM for the Navier-Stokes equations and these 8 modes capture more than 99.95\% energy for all Reynolds numbers included in the training. Figure~\ref{fig:eigen_history} shows the eigenvalues for 2D Navier-Stokes equations with $\gamma=0.1$. We use two different amplitudes for Taylor-green vortices (i.e., $\gamma=0.01$ and $\gamma=0.0$) to demonstrate the effectiveness of HAM framework in modeling different magnitudes of hidden physics. First we present the results for $\gamma=0.01$ which represents a comparatively easier case due to its small magnitude compared to the actual vorticity field. Figure~\ref{fig:ns_modes_500} and ~\ref{fig:ns_modes_1000} show the trajectories of the vorticity modal coefficients obtained by true projection, GP, and HAM framework. The GP deviates from the true projection modal coefficients due to the presence of source term which is not embedded in the GP model. The HAM framework can accurately predict the effect of source term though LSTM network at every time step, and we observe that the hybrid modal coefficients follow almost the same trajectory as true modal coefficients. We notice a small discrepancy between true and hybrid modal coefficients, especially for later modes (i.e., $a_7$ and $a_8$), as seen in Figure~\ref{fig:ns_modes_500} for $\text{Re}=500$. Figure~\ref{fig:ns_field} displays the vorticity field at the final time $t=20$ for FOM and the reconstructed field using true, GP, and hybrid modal coefficients. The GP fails to capture the correct orientation of two counter-rotating vortices at the final time for both $\text{Re}=500$ and $1000$. The HAM framework predicted the correct orientation of vortices with sufficient accuracy in comparison to the FOM vorticity field.

\begin{figure}
\centering
\mbox{
	\subfigure{\includegraphics[width=0.95\textwidth]{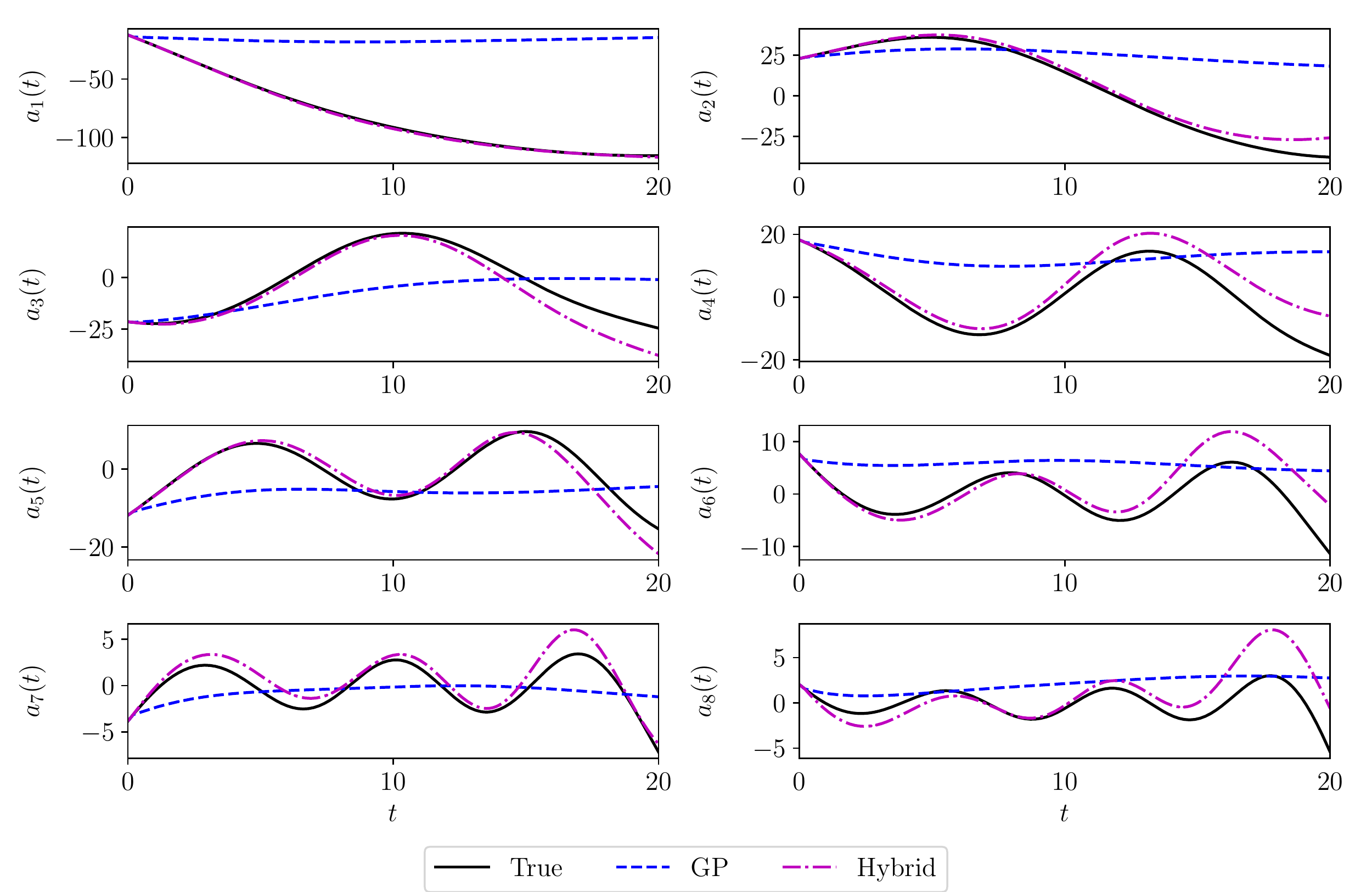}}
}
\caption{Temporal evolution of vorticity modal coefficients for the vortex-merger problem at $\text{Re}=500$ and $\gamma=0.1$.} 
\label{fig:ns_s2_modes_500}
\end{figure}

\begin{figure}
\centering
\mbox{
	\subfigure{\includegraphics[width=0.95\textwidth]{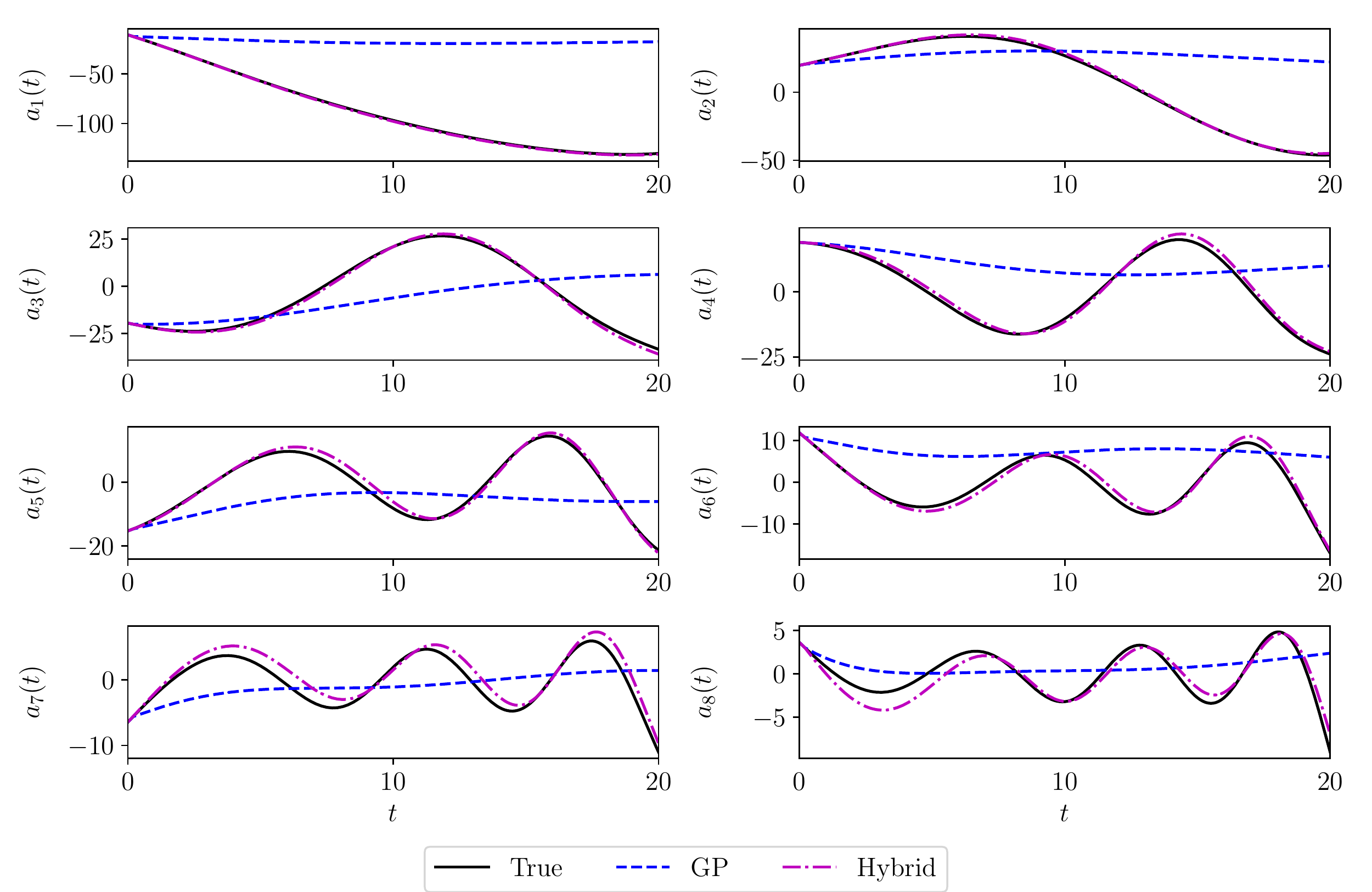}}
}
\caption{Temporal evolution of vorticity modal coefficients for the vortex-merger problem at $\text{Re}=1000$ and $\gamma=0.1$.} 
\label{fig:ns_s2_modes_1000}
\end{figure}

\begin{figure}
\centering
\mbox{
	\subfigure{\includegraphics[width=1.0\textwidth]{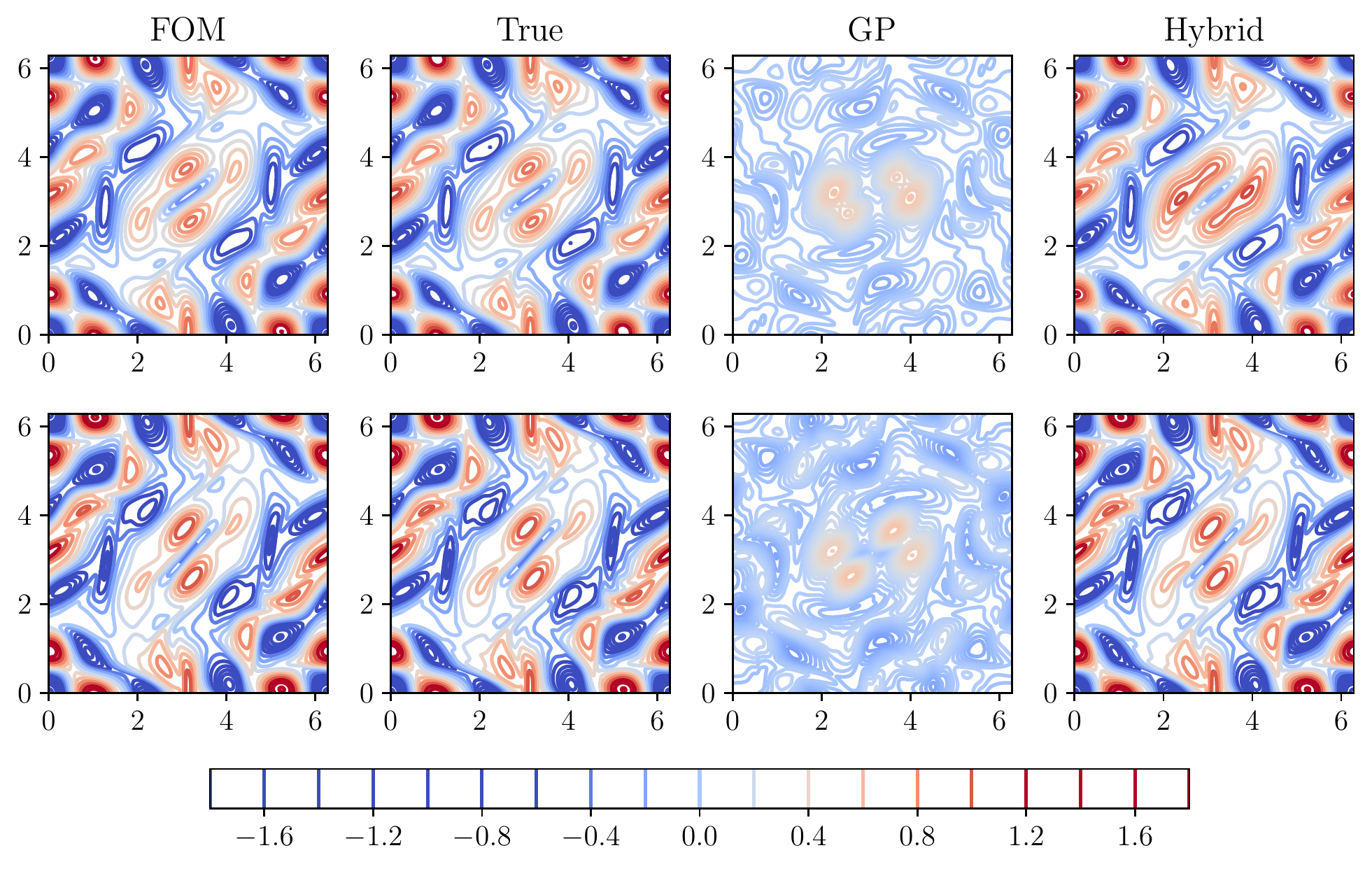}}
}
\caption{Vorticity field at time $t=20$ for the vortex-merger problem testing an interpolatory condition at $\text{Re}=500$ (top), and an extrapolatory condition at $\text{Re}=1000$ (bottom) with $\gamma=0.1$.} 
\label{fig:ns_s2_field}
\end{figure}

Next, we show the results of our numerical experiments with $\gamma=0.1$. Figure~\ref{fig:ns_s2_modes_500} and ~\ref{fig:ns_s2_modes_1000} shows the vorticity modal coefficients trajectories for $\text{Re}=500$ and $1000$. The GP fails to capture the correct dynamics as it does not incorporate any information about the hidden physics and for this test case, the major contribution comes from the unknown source term. The HAM framework, on the other hand is able to produce the correct dynamics for both $\text{Re}=500$ and $1000$ with little deviation near the final time for few modes. Figure~\ref{fig:ns_s2_field} shows the vorticty field at $t=20$ and it can be noticed that the HAM framework predicts the vorticity field with sufficient accuracy for both Reynolds numbers.  

\section{Discussion and conclusion} \label{sec:conclusion}
A hybrid ROM framework is presented for parameterized systems with hidden physics that uses GP model to compute modal coefficients and LSTM neural network to predict the unknown physics. The parameter governing the physical system is directly embedded in GP model and hence its effect on the unknown physics is considered implicitly. 

The accuracy of the HAM framework in predicting modal coefficients and reconstructed field is illustrated for two prototypical examples: Burgers and Navier-Stokes equations. The prediction for the interpolatory parameter by HAM framework is shown to be considerably more accurate than only physics-based GP model which cannot model the hidden physics. The evaluation of the HAM framework for extrapolatory parameter shows also a great agreement with the true projection, and yields results superior to the GP model. The numerical experiments with these two examples suggest that the hybrid modeling approaches have the potential to model a multiphysics system where there is a deviation in the physical model and the observed data.

We present the robustness of our approach using synthetic data sets generated either from an analytical solution or numerical simulation. Our approach can also be viewed as a data assimilation technique on a reduced dimensional space, which is a topic we intend to exploit further. The performance of the HAM framework in the presence of multiple source terms, noisy and realistic observed data will be investigated as a part of future studies.  

\section*{Acknowledgement}
This material is based upon work supported by the U.S. Department of Energy, Office of Science, Office of Advanced Scientific Computing Research under Award Number DE-SC0019290. O.S. gratefully acknowledges their support. 

Disclaimer. This report was prepared as an account of work sponsored by an agency of the United States Government. Neither the United States Government nor any agency thereof, nor any of their employees, makes any warranty, express or implied, or assumes any legal liability or responsibility for the accuracy, completeness, or usefulness of any information, apparatus, product, or process disclosed, or represents that its use would not infringe privately owned rights. Reference herein to any specific commercial product, process, or service by trade name, trademark, manufacturer, or otherwise does not necessarily constitute or imply its endorsement, recommendation, or favoring by the United States Government or any agency thereof. The views and opinions of authors expressed herein do not necessarily state or reflect those of the United States Government or any agency thereof.

\bibliographystyle{unsrt} 
\bibliography{ref}   

\end{document}